\documentclass[12pt]{iopart}

\usepackage{graphicx}
\usepackage{lineno}
\begin{document}

\title[Minimal Magnetic Energy Theorem]{Minimal Magnetic Energy Theorem}

\author{M C N Fiolhais and C Provid\^encia}

\address{Departamento de F\'isica, Universidade de Coimbra, 3004-516 Coimbra, Portugal}
\ead{miguel.fiolhais@cern.ch and cp@fis.uc.pt}
\begin{abstract}
Thomson's theorem states that static charge distributions in conductors only exist at the conducting surfaces in an equipotential
configuration, yielding a minimal electrostatic energy. In this work we present a proof for this theorem based on the variational
principle. Furthermore, an analogue statement for magnetic systems is also introduced and proven: the stored
magnetic field energy reaches the minimum value for superficial current distributions so that the magnetic vector potential points in the same direction
as the surface current. This is the counterpart of Thomson's theorem for the magnetic field. The result agrees with the fact that currents in superconductors are confined near the surface and indicates that the distinction between superconductors and hypothetical perfect conductors is fictitious.

\end{abstract}

\maketitle

\section{Introduction}
According to Thomson's theorem \cite{jackson}, given a certain
number of conductors, each one with a given charge, the charges
distribute themselves on the conductor surfaces in order to minimize
the electrostatic energy. Thomson's result
has been applied several times in the literature. In
particular, in papers of this journal it was used `to find the density
of the induced surface charge in the classical case of a point
charge in front of an infinite planar conductor' \cite{donolato} and
how `the use of the minimum energy principle to explain static
distributions of electric charge provides interesting physical
insights' \cite{coimbra}. Other interesting applications of the
Thomson's theorem are also found in \cite{computacional,florida}. Even though the proof of the theorem may 
be found with great detail in \cite{coulson,panofsky,landau}, a different approach not found in the literature, 
based on the variational principle, is presented in this paper.

Moreover, the analogue situation for the magnetic field is also
introduced: \emph{The magnetic field energy is minimal for superficial currents
distributions so that the magnetic vector potential points in the same direction
as the surface current.} However, since energy  conservation is assumed, this result cannot 
be applied to ohmic conductors in which a steady current can only exist if there is an electric field. Thus, it can only be used in classical perfect conductors and superconductors. Furthermore, it must be stressed that this theorem is not incompatible with Greiner's and Ess\'en's result on the magnetic field energy maximization \cite{greiner,essen}. Their work concerns currents in wires that can move in space while this study regards fixed currents. We shall prove this statement and Thomson's theorem by using a variational principle for the electromagnetic field energy functional.

\section{The proof of the theorem}
\subsection{Thomson's theorem}
In equilibrium, the static energy functional for
the electric field in a system composed of a conducting region surrounded by vacuum, may be written as:
\begin{eqnarray}
 U &&= \int_{V} \left\{  \rho \phi - {\epsilon_0 \over 2}
\left (\vec{\nabla} \phi \right ) ^2 \right \} \textrm{d}V
    +\int_{S} \sigma \phi \textrm{d}S \nonumber \\
   &&=  \int_{V_{\rm{in}}} \left\{ \rho \phi - {\epsilon_0 \over 2}
\left (\vec{\nabla} \phi \right )^2 \right \} \textrm{d}V
    -\int_{V_{\rm{out}}} {\epsilon_0 \over 2} \left
(\vec{\nabla} \phi \right )^2 \textrm{d}V
    +\int_{S} \sigma \phi \textrm{d}S \nonumber\\
   &&=  \int_{V_{\rm{in}}} \left \{ \rho \phi - {\epsilon_0 \over 2}
\left [ \vec{\nabla} \cdot \left (\phi \vec{\nabla} \phi \right ) -
\phi \vec{\nabla}^2 \phi \right ] \right \} \textrm{d}V \nonumber \\&&
    -\int_{V_{\rm{out}}} {\epsilon_0 \over 2} \left \{ \vec{\nabla} \cdot \left
(\phi \vec{\nabla} \phi \right ) - \phi \vec{\nabla}^2 \phi \right \}
\textrm{d}V+\int_{S} \sigma \phi \textrm{d}S
\label{energia}
\end{eqnarray}
The volume of the conductor is represented by $V_{\rm{in}}$ , the outside
volume is $V_{\rm{out}}$ and $S$ is the conductor boundary surface. Inside the conductor, we consider the volume charge distribution $\rho$ and on 
the conductor's surface, the sufarce charge distribution $\sigma$.
After applying Green's Theorem, the energy functional becomes:
\begin{eqnarray} U = \int_{V_{\rm{in}}} \left \{ \rho \phi -
{\epsilon_0 \over 2} \left ( \phi \vec{\nabla}^2 \phi \right ) \right \}
\textrm{d}V
    -\int_{V_{\rm{out}}} {\epsilon_0 \over 2} \left (\phi \vec{\nabla}^2 \phi \right)
\textrm{d}V \nonumber\\
+\int_{S} \left \{ \sigma \phi - {\epsilon_0 \over 2} \phi \hat{n} \cdot \left(
\vec{\nabla}^+ \phi - \vec{\nabla}^- \phi \right )\right \} \textrm{d}S
\end{eqnarray}
where $\vec{\nabla}^+$ and $\vec{\nabla}^-$ are the gradient operators at the surface in the upper and lower limits, respectively. 
Since the total charge in the conductor is limited and constrained
to the conductor volume and surface, one must include this restriction by introducing a Lagrange
multiplier. Therefore, the infinitesimal variation of the energy reads:
\begin{eqnarray}
\delta U &=&  \int_{V_{\rm{in}}} \left \{ \delta \phi \left( \rho + \epsilon_0
\vec{\nabla}^2 \phi \right) + \delta \rho \left ( \phi - \lambda
\right ) \right \} \textrm{d}V
    +\int_{V_{\rm{out}}} \delta \phi \epsilon_0
\vec{\nabla}^2 \phi \textrm{d}V \nonumber \\
&&+ \int_{S} \left \{ \delta \phi \left[ \sigma + \epsilon_0 \hat{n} \cdot
\left( \vec{\nabla}^+ \phi - \vec{\nabla}^- \phi \right ) \right] +
\delta \sigma \left( \phi - \lambda \right) \right \} \textrm{d}S
\end{eqnarray}

From the energy minimization, the Euler-Lagrange equations are:
\begin{equation}
V_{\rm{in}} \left\{
\begin{array}{l}
\vec{\nabla}^2 \phi = -\frac{\rho}{\epsilon_0} \\
\phi = \lambda \\
\end{array} \right.
\label{potencial}
\end{equation}
\begin{equation}
S \,\,\,\,\left\{
\begin{array}{l}
\sigma = - \epsilon_0 \hat{n} \cdot
\left( \vec{\nabla}^+ \phi - \vec{\nabla}^- \phi \right ) \\
\phi = \lambda \\
\end{array} \right.
\label{surface}
\end{equation}
\begin{equation}
V_{\rm{out}}\textrm{:}\, \vec{\nabla}^2 \phi = 0 \label{laplace}
\end{equation}

From these equations, in the minimum energy state, the potential is
constant inside the conductor which means the electric charge
distribution is superficial in an equipotential configuration
proving Thomson's theorem.

\subsection{Minimal Magnetic Energy theorem}

The same procedure will be applied to the magnetic field.
The energy functional for the magnetic field with no
time dependencies is written as:
\begin{eqnarray}
U &=&  \int_{V} \left \{ \vec{j} \cdot \vec{A} - {1 \over 2\mu_0} \left
(\vec{\nabla} \times \vec{A} \right ) ^2 \right \} \textrm{d}V
    +\int_{S} \vec{k} \cdot \vec{A} \textrm{d}S \nonumber \\
  &=&  \int_{V_{\rm{in}}} \left \{ \vec{j} \cdot \vec{A} - {1 \over 2\mu_0} \left
(\vec{\nabla} \times \vec{A} \right ) ^2 \right \} \textrm{d}V
    -\int_{V_{\rm{out}}} {1 \over 2\mu_0} \left
(\vec{\nabla} \times \vec{A} \right ) ^2 \textrm{d}V
    +\int_{S} \vec{k} \cdot \vec{A} \textrm{d}S \nonumber\\
  &=&  \int_{V_{\rm{in}}} \left \{ \vec{j} \cdot \vec{A} - {1 \over 2 \mu_0} \left
( \vec{\nabla} \cdot \left [\vec{A} \times \left(\vec{\nabla} \times
\vec{A} \right )\right ] +\vec{A} \cdot \left [\vec{\nabla} \times
\left(\vec{\nabla} \times \vec{A} \right ) \right ]\right ) \right \}
\textrm{d}V \nonumber \\
&&-\int_{V_{\rm{out}}} {1 \over 2 \mu_0} \left \{ \vec{\nabla} \cdot \left
[\vec{A} \times \left(\vec{\nabla} \times \vec{A} \right )\right ]
+\vec{A} \cdot \left [\vec{\nabla} \times \left(\vec{\nabla} \times
\vec{A} \right ) \right ]\right \} \textrm{d}V \nonumber \\
&&+ \int_{S} \vec{k} \cdot \vec{A} \textrm{d}S
\end{eqnarray}
Again, the volume of the conductor portion is represented by
$V_{\rm{in}}$, the outside volume is $V_{\rm{out}}$ and $S$ is the conductor
boundary surface. Inside the conductor, we consider the volume current distribution $\vec{j}$ and on
the conductor's surface, the sufarce current distribution $\vec{k}$. 
Using Green's Theorem, the energy functional becomes:
\begin{eqnarray}
U &=&  \int_{V_{\rm{in}}} \left \{ \vec{j} \cdot \vec{A} - {1 \over 2 \mu_0} \left
( \vec{A} \cdot \left [\vec{\nabla} \times \left(\vec{\nabla} \times
\vec{A} \right ) \right ] \right ) \right \}
\textrm{d}V \nonumber \\
&&-\int_{V_{\rm{out}}} {1 \over 2 \mu_0} \left \{ \vec{A} \cdot \left
[\vec{\nabla} \times \left(\vec{\nabla} \times
\vec{A} \right ) \right ]\right \} \textrm{d}V \nonumber \\
&&+ \int_{S} \left \{ \vec{k} \cdot \vec{A} - {1 \over 2 \mu_0} \vec{A} \cdot \left[ \hat{n} \times \left(
\vec{\nabla}^+ \times \vec{A} - \vec{\nabla}^- \times \vec{A}
\right) \right] \right \} \textrm{d}S
\label{variational}
\end{eqnarray}
As in the electric field case, some constraints should be imposed. In
particular, due to charge conservation, the electric current must obey the continuity equation
both in volume and at the surface \cite{ieee}:
\begin{eqnarray}
 \vec{\nabla} \cdot \vec{j} &=& 0  \\
 \vec{\nabla}_S \cdot \vec{k} &=& 0
\end{eqnarray}
where $\vec{\nabla}_S$ is the surface gradient operator. Notice that constraints are local, not global ones. In other words,
the Lagrange multipliers are not constant but space functions.
Therefore, the infinitesimal variation of the magnetic energy reads:

\begin{eqnarray}
&&\delta U =  \int_{V_{\rm{in}}} \left \{ \delta \vec{A} \cdot \left( \vec{j} -
\frac{1}{\mu_0}\left [\vec{\nabla} \times \left(\vec{\nabla} \times
\vec{A} \right ) \right ] \right) + \delta \vec{j} \cdot \left(
\vec{A}-\vec{\nabla}\lambda \right) \right \} \textrm{d}V \nonumber \\
&&-\int_{V_{\rm{out}}} \delta \vec{A} \cdot \frac{1}{\mu_0}\left
[\vec{\nabla} \times \left(\vec{\nabla} \times
\vec{A} \right ) \right ] \textrm{d}V \nonumber \\&&
+ \int_{S} \left\{ \delta \vec{A} \cdot \left[ \vec{k} + \frac{1}{\mu_0}
\hat{n} \times \left( \vec{\nabla}^+ \times \vec{A} - \vec{\nabla}^-
\times \vec{A} \right ) \right] 
+ \delta \vec{k} \cdot \left( \vec{A} - \vec{\nabla}_S \lambda
\right) \right \} \textrm{d}S
\end{eqnarray}
Hence, the Euler-Lagrange equations for this energy functional are:

\begin{equation}
V_{\rm{in}} \left\{
\begin{array}{l}
\vec{\nabla} \times \left(\vec{\nabla}\times
\vec{A}\right)=\vec{\nabla} \times \vec{B}=\mu_0
\vec{j} \\
\vec{A} = \vec{\nabla}\lambda \\
\end{array} \right.
\label{potencial2}
\end{equation}
\begin{equation}
S \,\,\,\,\left\{
\begin{array}{l}
\vec{k} =   \frac{1}{\mu_0} \hat{n} \times
\left( \vec{\nabla}^+ \times \vec{A} - \vec{\nabla}^- \times \vec{A} \right ) \\
\vec{A} = \vec{\nabla}_S\lambda \\
\end{array} \right.
\label{surface2}
\end{equation}
\begin{equation}
V_{\rm{out}}\textrm{:}\, \vec{\nabla} \times \left(\vec{\nabla}\times
\vec{A}\right)=\vec{\nabla} \times \vec{B} = 0
\label{laplace_mag}
\end{equation}

From eq. (\ref{potencial2}), the magnetic field must be zero inside the conductor, discarding any currents in volume ($\vec{j}=0$) in the
minimum energy state. On the other hand, from eq. (\ref{surface2}), one concludes that 
the potential vector $\vec{A}$ is tangent to the conducting surface. Since the magnetic field is zero in the inner volume, considering a cartesian coordinate system for a given surface element, if one of the axis is orthogonal to the surface element, the surface current becomes:
\begin{equation}
\vec{k} =  \frac{1}{\mu_0} \hat{n} \times \left[ \left( \vec{\nabla}_{\hat{n}} +   \vec{\nabla}_{S} + \vec{\nabla}_{S'} \right ) \times  \vec{\nabla}_S \lambda \right ] =   \frac{1}{\mu_0} \hat{n} \times \left[  \vec{\nabla}_{\hat{n}}  \times  \vec{\nabla}_S  \lambda  \right ]\\
\label{current}
\end{equation}
where $ \vec{\nabla}_{S}$ points in the same direction of $\vec{A} = \vec{\nabla}_S \lambda$ by definition and $ \vec{\nabla}_{S'}$ is orthogonal. From this result, it is clear that the surface current $\vec{k}$ will point in the same direction of $\vec{A}$. Notice that no restriction was imposed on the vector potential. Therefore this result is independent of the chosen gauge.
Even though the proof of the theorem does not include the presence of an external magnetic field, the result can be easily generalized. The presence of an external magnetic field demands another surface integral in eq (\ref{variational}) for currents at infinity. It does not affect the final result, however.

\section{Applications}

\subsection{Coaxial cable}

The Minimal Magnetic Energy Theorem can be illustrated in a simple
system such as the coaxial cable. A coaxial cable can
be modeled by an outer cylindrical conducting surface with radius
$b$, carrying an electric current $I$, and a massive cylindrical
conductor with 
radius $a$, carrying the same electric current in the opposite
direction. Furthermore, for this case study, the inner current is divided into a superficial current $I_1$ at $r=a$ and a current $I_2$ in the inner 
volume, so that $I=I_1+I_2$. The magnetic field is readily obtained using Amp\`ere's Law:

\begin{equation}
\vec{B} = \left\{
\begin{array}{ll}
\frac{\mu_{0}I_2}{2 \pi r} \frac{r^2}{a^2} \hat{\phi}& 0<r<a \\
\frac{\mu_{0}I}{2 \pi r} \hat{\phi}& a<r<b \\
0 & b<r \\
\end{array} \right.
\label{surface3}
\end{equation}

Thus, the magnetic field energy on a length $L$ of the cable is
given by:

\begin{eqnarray}
U & = & \frac{1}{2\mu_0} \int_{V} B^2 \textrm{d}V \nonumber\\
 & = & \frac{1}{2\mu_0} \int_{0}^a \left ( \frac{\mu_{0}I_2}{2 \pi r} \frac{r^2}{a^2} \right )^2 2 L \pi r \textrm{d} r + 
\frac{1}{2\mu_0} \int_{a}^b \left ( \frac{\mu_{0}I}{2 \pi r} \right )^2 2 L \pi r \textrm{d} r \nonumber \\
 & = & \frac{I_{2}^2 L \mu_0}{16 \pi}  + \frac{I^2 L \mu_0}{4 \pi} \log{b/a}
\end{eqnarray}

The magnetic energy reaches the minimum for $I_2=0$ and therefore, for $I=I_1$. In other words, the
magnetic energy attains its minimum value when the current only flows at the surface.
Furthermore, the current density was considered to be homogeneous in
the azimuthal coordinate, which is intuitively the minimum energy
configuration, leading to a tangent magnetic vector potential. A similar analysis for an electrostatic distribution can be found in \cite{coimbra}.

\subsection{Superconducting cylinder in a constant magnetic field}

\begin{figure}
\begin{center}
\includegraphics[width=9.cm]{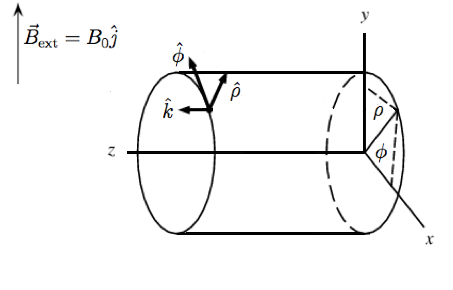}
\caption{Infinite superconducting cylinder in a constant magnetic field}
\end{center}
\end{figure}

The Minimal Magnetic Energy Theorem can also be useful to determine the configuration of the magnetic field and surface current distribution of a superconductor in equilibrium, for example, a cylindrical perfect conductor/superconductor with radius $R$ immersed in a external constant magnetic field (see Figure 1). In order to achieve the magnetic vector potential solution, one must solve the following differential equations:

\begin{equation}
\vec{\nabla} \times \vec{B} = \vec{\nabla} \times \left ( \vec{\nabla} \times \vec{A} \right )= 0 \\
\label{curl}
\end{equation}

However, instead of starting solving immideatly, one should look into the symmetries of the system. For instance, if the external constant magnetic field points in the same direction of the y-axis  and if the cylinder is considered to be infinite, there won't be any dependencies on the z coordinate (cylinder axis). Therefore, the magnetic field simplifies to:

\begin{equation}
\vec{B} = \hat{\rho} \frac{1}{\rho} \frac{\partial A_z}{\partial \phi} - \hat{\phi} \frac{\partial A_z}{\partial \rho} + \hat{k} \frac{1}{\rho} \left ( \frac{\partial  }{\partial \rho} (\rho A_{\phi}) - \frac{\partial A_{\rho}}{\partial \phi} \right )
\label{magnetic_2}
\end{equation}

Moreover, due to the symmetry of the system, the z-component of the magnetic field must be zero:
\begin{equation}
\frac{\partial  }{\partial \rho} (\rho A_{\phi}) - \frac{\partial A_{\rho}}{\partial \phi} = 0
\label{equacao}
\end{equation}

These assumptions and constraints transform eq. (\ref{curl}) into:

\begin{equation}
\vec{\nabla} \times \vec{B} = - \hat{k} \left ( \frac{1}{\rho^2} \frac{\partial^2 A_z}{\partial \phi^2} +  \frac{\partial^2 A_z}{\partial \rho^2} +  \frac{1}{\rho}  \frac{\partial A_z}{\partial \rho}   \right )= 0
\end{equation}
which is the well known Laplace equation. Before writing down the general solution, let's analyze the boundary conditions. At $\rho \rightarrow \infty$, the magnetic field must be equal to the external one: $\vec{B}_{\rm{ext}} = B_0 \hat{j} = B_0 (\hat{\phi} \cos \phi +\hat{\rho} \sin \phi ) $. Furthermore, since the magnetic field is zero inside the perfect conductor, from eq. (\ref{magnetic_2}) one concludes the magnetic potential vector must be constant inside the perfect conductor/superconductor. Therefore, the solution for this case study is then: 
\begin{equation}
A_z = \textrm{const.} + B_0 \left (  \frac{R^2}{\rho} - \rho \right ) \cos \phi \\
\end{equation}
Even though this is only the solution for the z-component of $\vec{A}$, the other components are not relevant to compute the magnetic field. However, the two remaining solutions can always be determined assuming the Coulomb gauge and using eq. (\ref{equacao}) while the boundary condition comes from eq. (\ref{current}) since the surface current must point in the same direction of the potential vector. Now everything is ready to calculate the magnetic field outside the cylinder:

\begin{equation}
\vec{B} = \hat{\rho} B_0 \left ( 1 - \frac{R^2}{\rho^2}  \right)\sin \phi  + \hat{\phi} B_0 \left( 1 + \frac{R^2}{\rho^2} \right )\cos \phi
\label{magnetic}
\end{equation}
which implies that,
\begin{equation}
\vec{B}(\rho=R) = 2 B_0 \cos \phi \,\, \hat{\phi}
\label{magnetic8}
\end{equation}
The knowledge of the magnetic field on the cylinder's surface allows determination of the surface current:
\begin{equation}
\vec{k} = 2 \frac{B_0}{\mu_0}  \cos \phi \,\, \hat{k}
\label{surfacecurrent}
\end{equation}
The total current obtained through integration of the surface current is zero as expected, otherwise the energy would diverge. A more detailed analysis on this problem can be found in \cite{cylinder}.

\subsection{Superconducting sphere in a constant magnetic field}
The same calculations can be performed for a superconducting sphere with radius $R$ in a constant magnetic field pointing in the z-axis direction (see Figure 2).

\begin{figure}[h]
\begin{center}
\includegraphics[width=7.cm]{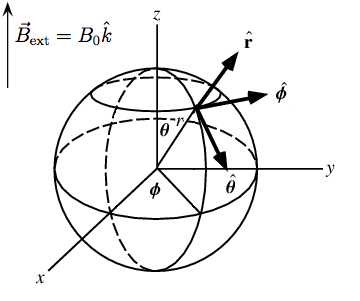}
\caption{Superconducting sphere in a constant magnetic field}
\end{center}
\end{figure}
Similarly to the cylinder case, eq. (\ref{curl}) is simplified a lot by implementing the symmetries of the system. Since the external constant magnetic field points in the same direction of the z-axis, there won't be any dependencies in the $\phi$ coordinate and the magnetic field along this coordinate must be zero. Therefore, the magnetic field simplifies to:

\begin{equation}
\vec{B} =\frac{1}{r \sin \theta} \frac{\partial} {\partial \theta} \left( A_\phi \sin \theta \right) \hat{r} -  \frac{1}{r} \frac{\partial }{\partial r}  \left( A_\phi r \right) \hat{\theta}
\label{magnetic_3}
\end{equation}
Again, these assumptions and constraints transform eq. (\ref{curl}) into:
\begin{equation}
\vec{\nabla} \times \vec{B} = - \hat{\phi} \frac{1}{r} \left [ \frac{\partial }{\partial r} \left ( \frac{\partial }{\partial r}  \left ( rA_\phi \right ) \right ) + \frac{1}{r} \frac{\partial }{\partial \theta} \left ( \frac{1}{\sin \theta} \frac{\partial }{\partial \theta}  \left ( A_\phi \sin \theta \right ) \right ) \right ] = 0
\label{curl_2}
\end{equation}
Since the magnetic field is zero inside the sphere, from eq (\ref{magnetic_3}), the magnetic vector potential must have the following form: 
\begin{equation}
A_\phi (r < R) = \frac{C}{r \sin \theta} 
\end{equation}
where C is a constant. In order to prevent the magnetic vector potential from diverging at $r=0$ and $\theta = 0$, the constant C must be zero. Furthermore, at $r \rightarrow \infty$, the magnetic field must be equal to the external one: $\vec{B}_{\rm{ext}} = B_0 \hat{k} = B_0 (\hat{r} \cos \theta -\hat{\theta} \sin \theta ) $. Therefore, the solution of  eq (\ref{curl_2}) for this case is:
\begin{equation}
A_\phi (r > R) = \frac{B_0}{2} \left (  r - \frac{R^3}{r^2} \right ) \sin \theta
\end{equation}
which leads to the following magnetic field outside the sphere,
\begin{equation}
\vec{B} = \hat{r} B_0  \left ( 1 - \frac{R^3}{r^3}  \right) \cos \theta  - \hat{\theta} B_0 \left( 1 + \frac{1}{2} \frac{R^3}{r^3} \right )\sin \theta
\label{magnetic_4}
\end{equation}
From this result, the magnetic field at the sphere surface becomes:
\begin{equation}
\vec{B} = - \frac{3}{2} B_0  \sin \theta  \,\, \hat{\theta}
\label{magnetic_5}
\end{equation}
Using eq (\ref{surface2}), the surface current density is therefore:
\begin{equation}
\vec{k} = - \frac{3}{2} \frac{B_0}{\mu_0} \sin \theta  \,\, \hat{\phi}
\label{current_3}
\end{equation}
Unlike the infinite superconducting cylinder in a constant magnetic field, the superconducting sphere has a total non-zero electric current, $I = 3 \frac{B_0}{\mu_0}R $. This current is essential in order to keep the magnetic field from entering inside the spherical superconductor. A similar approach to this problem can also be found in \cite{sphere}.

\section{Conclusions}
In conclusion, the energy approach to the nature of magnetic fields has provided several interesting physical results such as \cite{essen2,essen3,essen4,essen5,essen6}. In particular our theorem, analogously to Thomson's theorem in electrostatics, allows the prediction of surface current distributions in superconductors at equilibrium. Furthermore, it may change the distinction between superconductors and perfect conductors.

In the pedagogical literature on superconductivity a distinction between superconductors and hypothetical perfect conductors is sometimes made. Both have zero resistivity below some critical temperature but when cooled below this temperature, while penetrated by constant external magnetic fields, the superconductor expels the field (the experimentally observed Meissner effect) but, in the corresponding thought experiment, the perfect conductor does not (see e.g. Poole et al. [19]). From the Minimal Magnetic Energy Theorem, one concludes that energy minimization and expelling external fields are the same thing. Therefore, the perfect conductors would in fact behave like superconductors in their approach to thermodynamic equilibrium.

\end{document}